\documentclass[aps,prb,preprint,preprintnumbers,amsmath,amssymb,superscriptaddress,fleqn]{revtex4-1}

\usepackage{graphicx}
\usepackage{bm}
\usepackage{color}

\usepackage[T1]{fontenc}
\usepackage[cp1250]{inputenc}

\usepackage[firstpage]{draftwatermark}
\usepackage[breaklinks]{hyperref}
\usepackage[all]{hypcap}

\hypersetup{
    plainpages=false,
    bookmarks=false,         
    unicode=false,          
    pdfborder={0 0 0}
    pdftoolbar=true,        
    pdfmenubar=true,        
    pdffitwindow=false,     
    pdfstartview={FitH},    
    pdftitle={LNCMI Annual Report 2012},    
    pdfauthor={Duncan Maude},     
    pdfproducer={LNCMI-CNRS-UJF-UPS-INSA}, 
    pdfkeywords={High} {Magnetic} {Fields}, 
    pdfnewwindow=true,      
    linktoc=section,
    colorlinks=true,       
    linkcolor=blue,          
    citecolor=red,        
    filecolor=magenta,      
    urlcolor=blue           
}

\begin{document}

\title{Intervalley scattering of excitons and trions in monolayer WSe$_{2}$ \\under strong excitation}

\author{A. A. Mitioglu}
\affiliation{LNCMI, CNRS-UJF-UPS-INSA, Grenoble and Toulouse, France} \affiliation{Institute of Applied Physics,
Academiei Str. 5, Chisinau, MD-2028, Republic of Moldova}

\author{Ł. Kłopotowski}
\affiliation{Institute of Physics, Polish Academy of Sciences, al. Lotników 32/46, 02-668 Warsaw, Poland}

\author{D.~K.~Maude}
\affiliation{LNCMI, CNRS-UJF-UPS-INSA, Grenoble and Toulouse, France}

\author{G. Deligeorgis}
\affiliation{FORTH-IESL, Microelectronics Research Group, P.O. Box 1527, 71110 Heraklion, Crete, Greece}

\author{S. Anghel}
\affiliation{Institute of Applied Physics, Academiei Str. 5, Chisinau, MD-2028, Republic of Moldova}
\affiliation{Ruhr-Universit\"at Bochum, Anorganische Chemie III, D-44801 Bochum Germany}

\author{L. Kulyuk}
\affiliation{Institute of Applied Physics, Academiei Str. 5, Chisinau, MD-2028, Republic of Moldova}

\author{P. Plochocka}
\email{paulina.plochocka@lncmi.cnrs.fr}
\affiliation{LNCMI, CNRS-UJF-UPS-INSA, Grenoble and Toulouse, France}

\date{\today}

\begin{abstract}
Monolayer transition metal dichalcogenides offer the possibility of optical control of the valley degree of freedom. In order to asses the potential of these materials in applications, detailed knowledge of the valley dynamics is essential. In this work, we apply low temperature time-resolved photoluminescence (PL) measurements to investigate exciton valley relaxation dynamics and, in particular, its behavior under strong excitation. At the lowest excitation powers the inter valley scattering time is $\simeq 50$ ps, but shortens by more than a factor of two at the highest powers. We attribute this acceleration to either heating of the exciton system or the presence of a dense exciton gas, which could influence the exciton valley properties. Furthermore, we analyze the PL dynamics of excitons and trions. We find that the PL decays for all peaks are bi-exponential and approximately independent of the excitation power. We attribute the short decay to radiative recombination and escape to a reservoir of dark states. The long decay is ascribed to a transfer of excitons back from the reservoir. For the first time, we evaluate the exciton PL decay time of $\simeq$ 10 ps. The latter process is valley-conserving and occurs on a timescale of $\simeq$ 50 ps
\end{abstract}

\maketitle


Since their isolation, atomically thin two dimensional (2D) semiconductors have attracted great interest due to their remarkable electronic properties and possible applications \cite{Novoselov05p,Butler13,Xu13a}. In particular, monolayer transition
metal dichalgogenides (TMDs) show significant potential in electronic \cite{Radisavljevic11,Lopez13,Krasnozhon14} and optoelectronic \cite{Ross14,Ubrig14,Zhang14} applications. Unlike bulk TMDs, monolayers have a direct band gap located at two non-equivalent valleys $(\pm K )$ at the corners of a hexagonal Brillouin zone.~\cite{Mak10,Splendiani10} However, perhaps the most fascinating properties of monolayer TMDs result from the lack of inversion symmetry and strong spin-orbit coupling. The electron spin and valley degrees of freedom are coupled and the resulting optical selection rules are valley contrasting; interband transitions in $\sigma^{\pm}$ polarizations involve electron-hole pairs (excitons) in the $\pm K$ valleys, respectively and thus the valley polarization can be monitored through the
circular dichroism.\cite{Xiao12,Xu14} Polarization resolved steady-state photoluminescence (PL) measurements reveal a high valley polarization for monolayer MoS$_{2}$~\cite{Cao12,Mak12,Zeng12} or WSe$_{2}$\cite{Jones13} suggesting a
valley relaxation time comparable to the exciton lifetime. On the other hand, time-resolved PL studies showed that the latter is rather short -- of the order of 10~ps,\cite{Korn11,Yan14,Lagarde14,Wang14} setting a limit on the valley
polarization lifetime.

The observation of a robust valley polarization opens the way for applications in novel {\em valleytronic} devices, where the valley degree of freedom is manipulated. However, to exploit the potential of TMDs in this respect, a detailed understanding of the exciton PL dynamics and valley/spin relaxation processes is crucial. To this end, several time-resolved techniques have been  recently applied to study the exciton dynamics in TMDs. In particular, polarization resolved pump-probe spectroscopy suggests that the decay of the circular dichroism occurs on a timescale in the range
of a few to almost 100~ps.\cite{Wang13,Mai14,Mai14a,Zhu14} The proposed mechanism underlying the loss of valley polarization is long range electron-hole exchange interaction, which couples the bright exciton states.\cite{Mai14a,Yu14,Glazov14}


In this paper, we investigate the exciton dynamics and inter-valley exciton relaxation in monolayer WSe$_{2}$ via time- and polarization-resolved PL measurements. In particular, we discuss the excitation density dependence of the recombination processes and of the valley polarization dynamics. We show that with increasing excitation power, changes in the PL decay are controlled by the saturation of non-radiative trap states. This is accompanied by a strong increase in the inter-valley scattering rate. We demonstrate that this acceleration affects only cold, thermalized excitons and propose that it is due to an increased exciton temperature or to a modification of valley properties of a dense exciton gas.

Single bulk crystals of 2H-WSe$_2$ (the hexagonal 2H-polytype of tungsten diselenide) were grown using chemical vapor transport with bromine as the transport agent. Monolayer flakes of WSe$_{2}$ were obtained from these crystals using mechanical exfoliation. For the PL measurements, the sample was placed in a helium flow cryostat with optical access. Since the size of monolayer flakes is on the order of a few $\mu$m, the PL measurements were performed with excitation and collection using a microscope objective with a numerical aperture NA = 0.66 and magnification of $50\times$ in a back scattering configuration. The typical diameter of the laser spot was approximately 1~$\mu$m. Steady-state $\mu$-PL signal was excited with a 640 nm laser and the spectra were recorded using a spectrometer equipped with a liquid
nitrogen cooled CCD camera. The time-resolved $\mu$-PL signal was excited with the frequency doubled tunable output of an optical parametric oscillator (OPO), synchronously pumped with a mode-locked Ti:Sapphire laser. The typical temporal
pulse width was $\simeq 300$~fs with a repetition rate of 80~MHz. The time resolved PL signal was dispersed by an imagining spectrometer and detected using a synchroscan streak camera with a temporal resolution of 3~ps. Circularly polarized excitation and detection were employed to create and probe the non-equilibrium valley occupation.

\begin{figure}
\begin{center}
\includegraphics[width=0.7\textwidth]{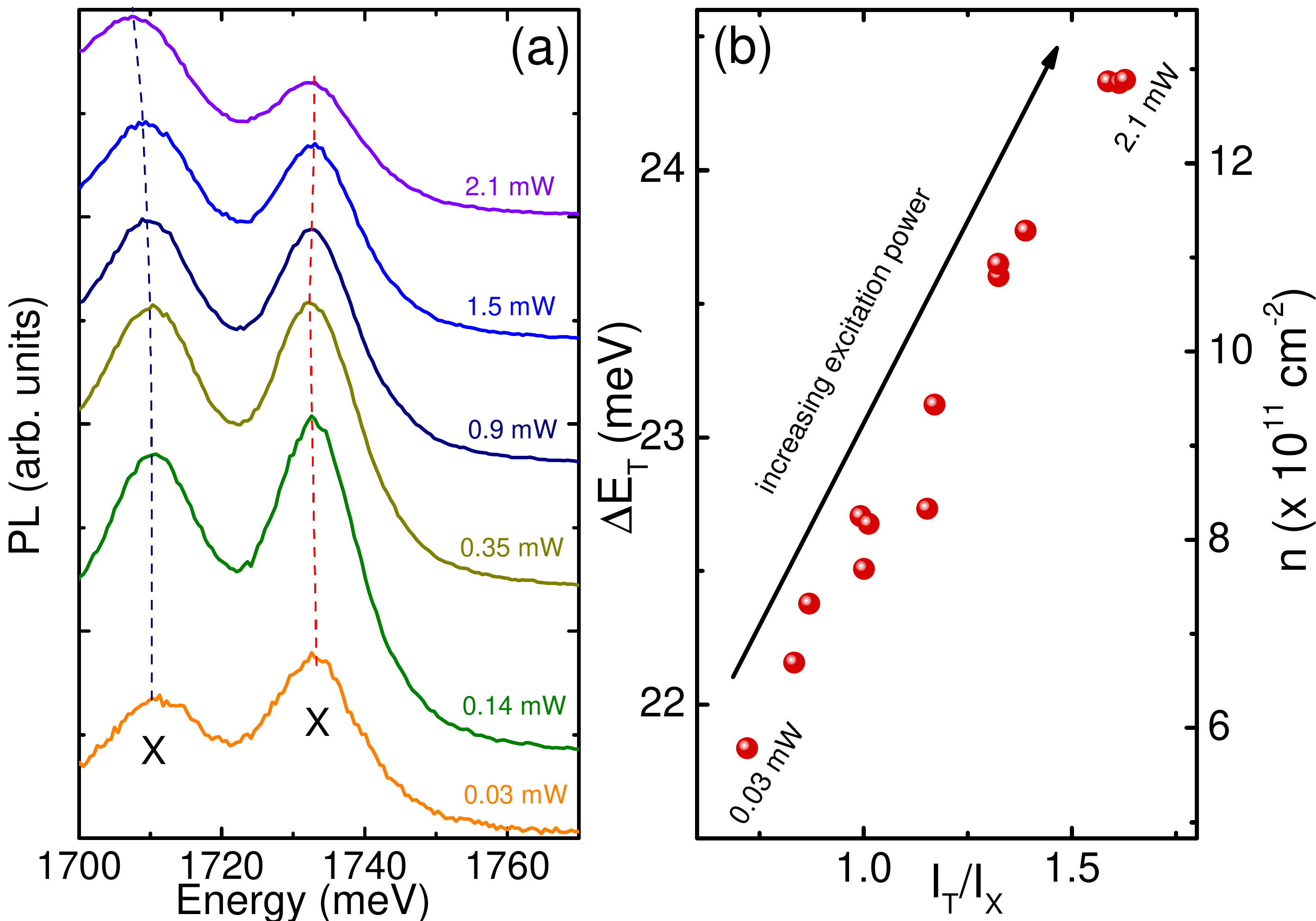}
\end{center}
\caption{ (a) Power dependence of the cw PL spectra of a WSe$_2$ monolayer measured at $T = 4$ K. Dashed lines are
guides to eye indicating increasing energy separation $\Delta E_T$ between the exciton and trion PL peaks with
increasing excitation power. (b) The dependence of $\Delta E_T$ on exciton-trion PL intensity ratio. $\Delta E_T$ is
converted to electron concentration in the right axis. See text for details.}\label{PL}
\end{figure}

We start with a discussion of the steady-state PL of excitons and
trions. Typical spectra obtained at $T=4$~K from a WSe$_2$ monolayer
as a function of excitation power are presented in Fig.~\ref{PL}(a).
The peak centered at $\sim$ 1733 meV corresponds to the
recombination of the A-exciton, as previously reported for WSe$_2$
monolayers.\cite{Zeng13,Zhao13,Jones13,Wang14,Mitioglu15} The peak,
appearing at $\sim$ 1710 meV is the recombination of the negatively
charged exciton (trion). At lower energies (not shown here) a series
of peaks corresponding to localized exciton states is
seen.\cite{Wang14,Zhu14} With increasing excitation power, the trion
emission gains in intensity with respect to the exciton emission.
Simultaneously, the energetic separation between the exciton and
trion peaks, $\Delta E_T$, increases. In Fig.~\ref{PL}(b), we plot
$\Delta E_T$ as a function of the intensity ratio of the trion and
exciton emission $I_T/I_X$. We can write $\Delta E_T = E_D + E_F$,
where $E_D$ and $E_F$ are the trion dissociation energy (the energy
required to unbind the electron from a neutral exciton) and Fermi
energy, respectively. The latter term results from the fact that
upon recombination of the trion, the electron occupies an available
state above the Fermi level.\cite{Huard00,Mak12} $\Delta E_T$
increases roughly linearly with excitation power, reflecting the
increase of $E_F$, \emph{i. e.}, the increase of the number of free
carriers. The plot of $\Delta E_T$ as a function of $I_T/I_X$
demonstrates that the ratio is directly proportional to the number
of excess carriers. Extrapolating $\Delta E$ to $I_T/I_X=0$ yields
the trion dissociation energy $E_D = 19.8 \pm 0.3$~meV, a value
typical for monolayer
WSe$_{2}$.\cite{Jones13,Wang14,Zhu14,Mitioglu15} We note that the
magnitude of $E_D$ depends not only on the TMD material, but also on
strain and dielectric environment.\cite{Mitioglu13} The power
dependence of $E_F$, shows that the number of photoexcited free
carrier can be as high as $1.2 \times 10^{12}$ cm$^{-2}$, for the
highest excitation power used of 2.1~mW.

Before turning to the time-resolved PL results, let us comment on the nature of the PL excitation. For steady-state experiments the excitation energy is $E_L = 1937$ meV, \emph{i. e.}, the excitons are created with an excess kinetic energy of $E_k = E_L - E_A \approx 200$~meV, where $E_A = 1733$ meV is the PL energy of the A exciton measured at low excitation power (see Fig.~\ref{PL}). The excitation is therefore below the WSe$_{2}$ band gap which is equal to about 2.5~eV.\cite{Ramasubramaniam12,Hanbicki15} Although quasi-resonant (well below the continuum), the excitation results in the creation of a large density of free carriers. This is due to the dynamic photo-ionization of neutral donor states present in our n-type samples.\cite{Mitioglu13}

\begin{figure}
\includegraphics[width=0.6\textwidth]{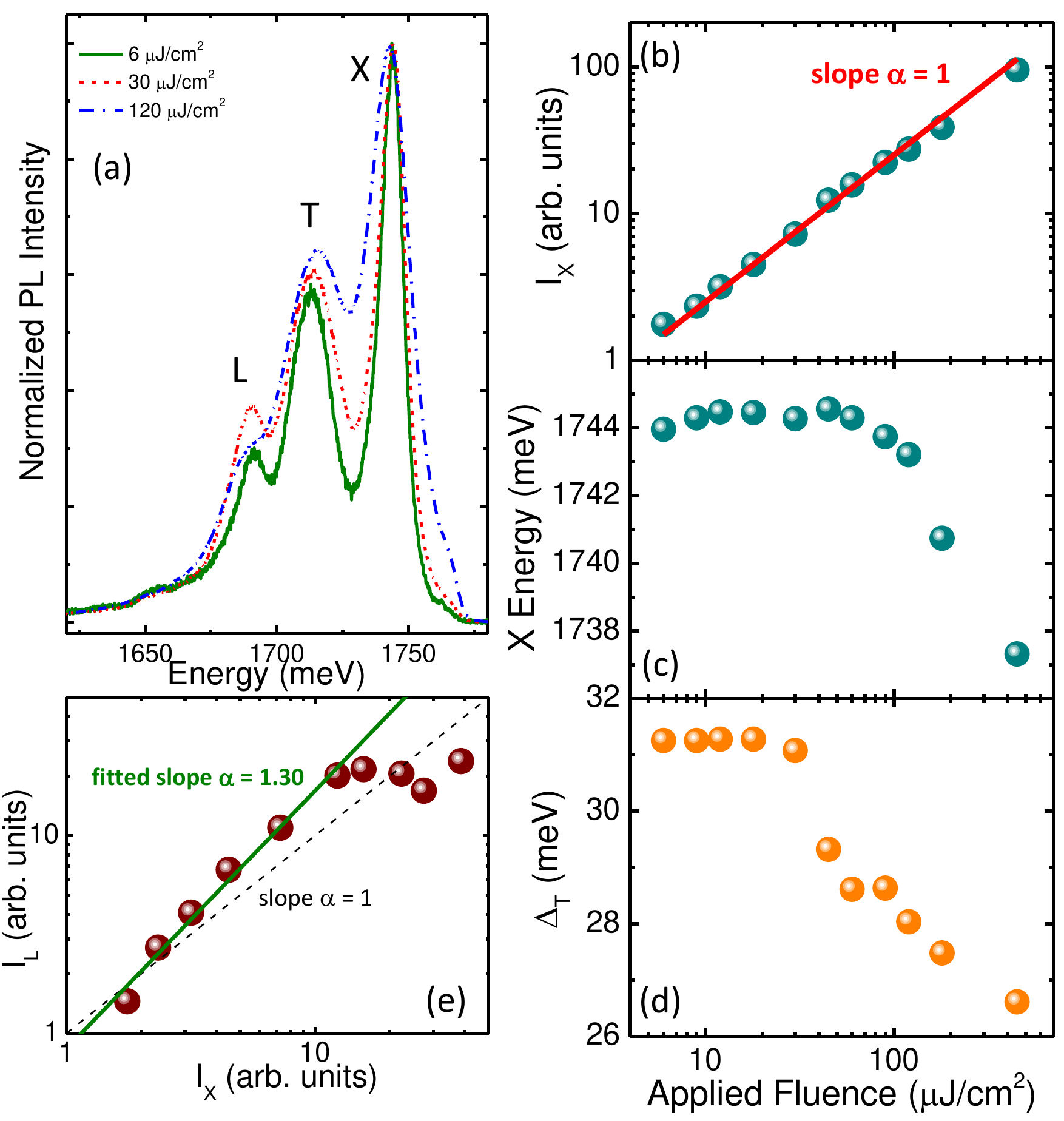}
\caption{ (a) Normalized PL spectra for pulsed excitation and various excitation fluences. (b) Fluence dependence of
the integrated intensity of the exciton (X) transition compared with a calculated linear dependence. (c) Fluence
dependence of the exciton transition energy. The redshift above 60 $\mu$J/cm$^{-2}$ indicates the onset of laser
heating. (d) Fluence dependence of $\Delta E_T$. (e) The dependence of the integrated intensity of the L transition on
the integrated intensity of the exciton transition. Fitted power law dependence yields an exponent of 1.3. For
comparison, a linear dependence is plotted.} \label{trpl}
\end{figure}

In Fig.~\ref{trpl}(a), we plot the normalized PL spectra excited with a pulsed laser at $E_L = 1850$~meV, which is close to resonance with an excited state of the A exciton.\cite{Wang15} In this case, $E_k \approx 120$ meV. The presented spectra are the sum of PL signals in $\sigma^+$ and $\sigma^-$ detection polarizations. Three peaks are observed: the exciton recombination (X), the trion recombination (T), and a third peak which we label as L. In Fig. \ref{trpl}(b), we show the dependence of the time-integrated PL intensity of the X peak on the excitation fluence. We find that the exciton intensity grows perfectly linearly over the whole fluence range used in the experiment. This allows us to conclude that under our experimental conditions, the exciton population does not exhibit any saturation from many body effects, such as biexciton formation\cite{You15} or exciton-exciton annihilation.\cite{Kumar14} This conclusions remains valid even at at the highest fluences where lattice heating induces a significant broadening and a redshift of the exciton emission (see Fig. \ref{trpl}(c)). The absence of the exciton-exciton annihilation suggests that this process is only efficient when assisted by exciton diffusion, which requires long exciton lifetimes and thus elevated temperatures.\cite{Mouri14}

The analysis of the PL spectra excited with OPO pulses reveals that, contrary to the steady state case, with increasing excitation power the trion intensity does not increase significantly compared to the exciton intensity, and $\Delta
E_T$ does not increase with excitation power -- see Fig. \ref{trpl}(d). This finding, corroborated on $\simeq10$ different flakes, shows that increasing the pulsed excitation power does not result in an increased number of free electrons. We therefore conclude, that already at the lowest useable excitation density, the pulse power is high enough to ionize all the available donors. In fact, $\Delta E_T$ decreases as the power is increased, even before the heating effects set in. We attribute this effect to a screening of the trion binding by the increasingly dense neutral exciton gas. The roughly linear increase of the trion peak with fluence (see Fig. \ref{trpl}(a)) reflects the creation of charged excitons from the carriers created by photoexcitation.

The origin of the L peak, separated by about 54 meV from the exciton peak, was previously assigned to either recombination of excitons bound on localized states \cite{Wang14,Zhu14,Wang15}, fine structure of the negative trion,\cite{Jones13} or very recently to the recombination of biexcitons.\cite{You15} As shown in Fig.~\ref{trpl}(e) the L line indeed exhibits a super linear increase with respect to the exciton at low fluences as expected for a biexciton. However, the intensity saturates at high excitation powers, which together with the absence of any deviation from the linear dependence of the exciton intensity on excitation power, precludes the assignment of the L line to biexciton recombination. The fine structure splitting of the trion resulting from electron-electron exchange interaction can also be excluded as the origin of the L emission. The recently calculated fine structure splitting of
$\simeq6$~meV,\cite{Yu14d} is much less than the separation between the T and L peaks, which is $\approx$~30 meV. Therefore, we assume that the L line originates from localized states. Further arguments in favor of this assignment are presented below.

\begin{figure}
\includegraphics[width=0.6\textwidth]{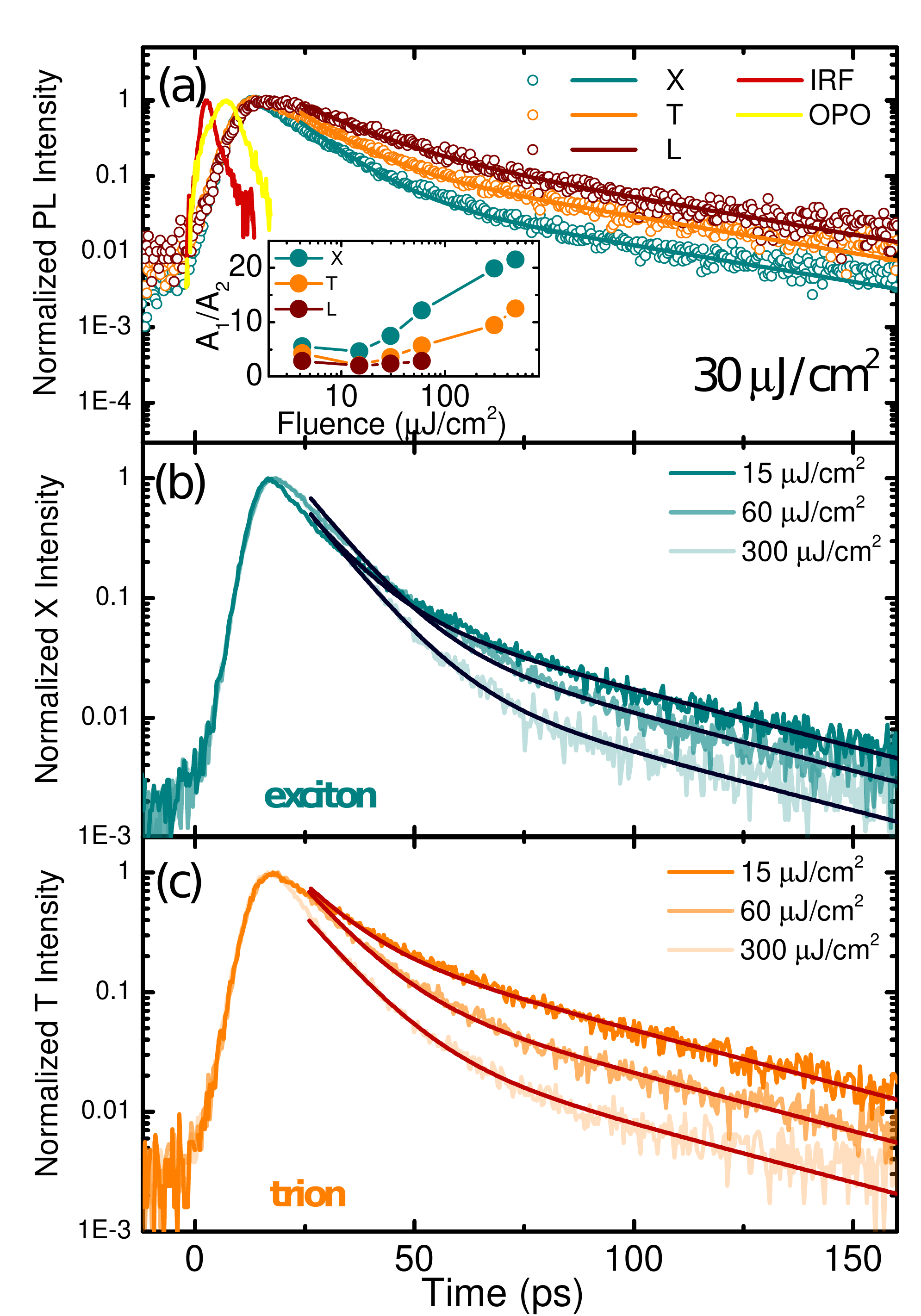}
\caption{ (a) Comparison between the decay traces for the exciton, trion and the L line. IRF denotes the instrument
response function. Its full width at half maximum is 3 ps. OPO denotes temporal trace of the excitation pulse. Inset:
the fluence dependence of the ratio between the fast and slow decay component (see text). (b) and (c) Temporal
evolution of the exciton and trion PL for various excitation fluences. Results of fitting of a bi-exponential decay are
plotted as lines.} \label{decays}
\end{figure}

In Fig.~\ref{decays}(a), we compare the decay dynamics of the exciton, trion, and the L emission and find that the exciton decay is the fastest and that of the L line is the slowest in agreement with a recent report.\cite{Wang14} Our temporal resolution does not allow us to resolve the rise time of the PL signal. All the temporal profiles can be reproduced by convoluting a Gaussian temporal profile of the OPO excitation pulse (characterized by a full width half maximum of about 7~ps) with a bi-exponential decay. This allows us to conclude that the thermalization of excitons occurs on a timescale shorter than the resolution of our measurement, \emph{i. e.} shorter than a few ps. The fluence dependence of the PL decay traces for the exciton and trion transitions are shown in Fig.~\ref{decays}(b) and ref{decays}(c), respectively.
The decays are fitted with $A_1 \exp(-t/t_1) + A_2 \exp(-t/t_2)$ and the fit is performed simultaneously to all seven decay traces for fluences ranging from 4.2 to 480 $\mu$J/cm$^2$. 
For the exciton and the trion, the fitting yields $t_1 = 9 \pm 1$ ps
and for the L transition $t_1 = 13 \pm 2$ ps. The short decay time is related to radiative recombination and/or transfer of excitons to the non-radiative states. Thus, in principle $t_1$ sets the lower limit for the intrinsic radiative recombination time, in agreement with other reports on WSe$_{2}$\cite{Wang14} and other TMDs.\cite{Korn11,Lagarde14} However, recent calculations show that the intrinsic exciton lifetime in WSe$_{2}$ at 4 K is equal to $\approx 4$~ps and increases sharply with temperature at a rate of $1-10$~ps/K.\cite{Palummo15} This suggests that the initial PL decay could be mostly due to radiative recombination. More insight into the origin of the $t_1$ time is provided below, in the discussion of the PL polarization.

The fitted value of $t_2 = 45 \pm 5$ ps of the long decay time constant is the same for all three transitions. We have measured several different flakes and $t_1$ always remains in the same range, while $t_2$ can be as high as 70~ps. We
interpret the long decay as originating from a transfer of excitons from a reservoir of non-radiative (dark) defect states. Thus, $t_2$ represents the exciton lifetime in the dark reservoir. As the excitation power is increased, the reservoir becomes progressively filled and, as a consequence, the ratio between the amplitude of the short and long decays, $A_1/A_2$ is increased for the exciton and the trion, as shown in the inset to Fig.~\ref{decays}(a). The observed saturation of the dark reservoir with increasing power shows that it is not related to states out of the
radiative cone, \emph{i. e.} with a high kinetic energy. For such states, with increasing excitation power, no saturation is expected; we should observe rather the opposite effect with the long decay amplitude increasing with respect to the short one.

We are unable to resolve any variation of $A_1/A_2$ for the L line.
However, the decays of the L line could only be measured up to 60 $\mu$J/cm$^2$. For higher fluences, the measurement
was precluded by the broadened trion line and decreased intensity of the L line (see Fig. \ref{trpl}(a)).

\begin{figure}
\includegraphics[width=0.5\textwidth]{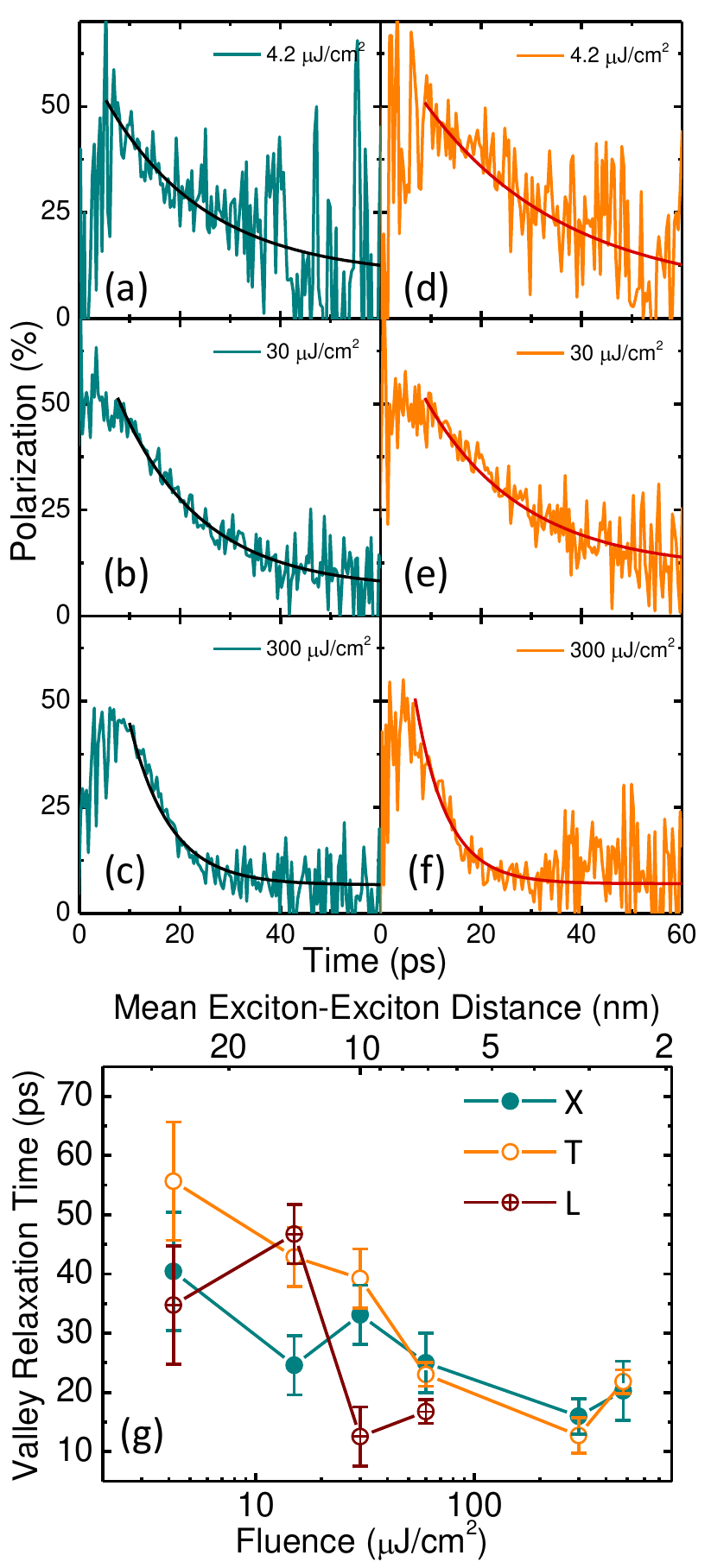}
\caption{ Temporal evolution of the circular polarization of the exciton PL (a)-(c) and trion PL (d)-(f) for various
excitation fluences. (g) Valley relaxation time ($t_{\text{iv}}$) extracted from the measured polarization decays,
plotted as a function of the excitation fluence, for the three PL peaks shown in Fig. \ref{trpl}(a).} \label{polar}
\end{figure}

We now turn to the discussion of the temporal dependence of the circular polarization of the PL signal. As shown by many authors, the circular polarization degree directly monitors the valley polarization of the excitons in monolayer
TMDs.\cite{Xiao12,Xu14,Cao12,Mak12,Zeng12,Lagarde14,Wang14,Wang13,Mai14,Mai14a,Zhu14,Kioseoglou12} We define the degree of circular polarization of the time-resolved PL signal as $\varrho = (I^--I^+)/(I^-+I^+)$, where $I^{\pm}$ are the intensities of the PL signal in $\sigma^{\pm}$ polarizations. The temporal decay of $\varrho$ reflects the decay of the valley polarization. In Fig.~\ref{polar}(a)-(f) we demonstrate temporal dependence of $\varrho$ of the exciton and trion emission. It shows that during the arrival of the excitation pulse $\varrho$ is approximately constant, with a value of about 50\%. It is important to note that this initial polarization value is nearly unaffected by increased excitation fluence. $\varrho$ starts to decay after a delay of $\simeq 10$~ps, \emph{i. e.} after the excitation pulse has vanished. $\varrho$ decays roughly exponentially to a non-zero value of about 5\%. This low polarization value and
the large noise level due to weak PL signal at time delays $>40$ ps make it impossible to conclude whether this polarization is subject to a long-lived decay\cite{Wang14} or rather constitutes an equilibrium polarization. Since its presence coincides temporally with the domination of the long decay of the PL signal, we suppose that this polarization results from a valley conserving transfer of excitons from the reservoir of dark states. Crucially, it is clearly seen from Fig.~\ref{polar} that as the excitation fluence is increased, the decay of $\varrho$ is accelerated. In the
following, we describe the currently accepted valley relaxation mechanism and within its framework discuss the possible origins of the observed shortening of the PL polarization lifetime.

Since the excitation is quasi-resonant and $\sigma^+$-polarized, due to the valley contrasting selection rules, the excitons are formed uniquely in the $+K$ valley.\cite{Xiao12,Xu14} However, since the excitons are created with a
non-zero center-of-mass momentum $k$ resulting from the excess energy $E_k$, an efficient depolarization process takes place. As proposed by many authors,\cite{Mai14,Yu14,Glazov14} the underlying mechanism is related to the long-range part of the electron-hole exchange interaction, which can be treated as an effective magnetic field $\mathbf{\Omega}$ around which the valley index precesses. Momentum scattering events lead to a randomization of the field direction and
thus the valley relaxation by the mechanism proposed for quantum well excitons.\cite{Maialle93} We assume a strong scattering regime, in which $\mathbf{\Omega} \tau_p \ll 1$, where $\tau_p$ is the momentum scattering time. Under such conditions the intervalley scattering time is $t_{\text{iv}} = (\langle \mathbf{\Omega}^2 \tau_p \rangle)^{-1}$, where the brackets denotes averaging over the distribution of exciton energies.\cite{Maialle93,Yu14,Glazov14} Since
$\mathbf{\Omega} \sim k$,\cite{Yu14} the valley polarization, and thus the circular polarization of PL, exhibit a strong dependence on the excitation energy.\cite{Kioseoglou12,Wang15}. In particular, for the values of $E_k$ from our
experiments, an initial depolarization time on the order of 10~fs is expected.\cite{Yu14} After thermalization and subsequent cooling, which occur during first $\simeq 0.5$~ps after excitation,\cite{Nie14} the average $k$ value of the
exciton population is reduced and, consequently, the valley depolarization slows down.

In order to reproduce the measured temporal dependence of $\varrho$ and to extract the experimental values of $t_{\text{iv}}$, we employ a simple rate equation model for the exciton populations $N^{\pm}$ in the $\pm K$ valleys:
\begin{equation}
\frac{d N^{\pm}}{dt} = - \frac{N^{\pm}}{t_{\text{rec}}} \mp \frac{N^+-N^-}{t_{\text{iv}}},
\label{rateeq}
\end{equation}
where $t_{\text{rec}}$ is the recombination time. Taking into account two exciton populations, neglects the fact that the exciton exhibits a richer fine structure with two additional dark configurations with parallel electron and hole spins. However, since the long-range electron-hole exchange couples only bright exciton states,\cite{Maialle93,Yu14} only these two states are involved in the inter-valley scattering process, which allows us to neglect the dark states.
In the case of the negatively charged trion, there are six bright spin/valley configurations split by both electron-hole and electron-electron exchange interactions.\cite{Jones13} Since the splitting between the ground and the
first excited trion configuration is about 6~meV,\cite{Yu14d} which is an order of magnitude larger than the thermal energy at the temperature of the experiments, we assume that only the ground configuration is populated and neglect the higher energy ones.

To account for the ultra fast initial depolarization, we shift the time origin by about 10~ps, \emph{i. e.}, to the time delay when the exciting OPO pulse has vanished and $\varrho$ starts to decay. Furthermore, we choose the initial conditions as $N^+(0) = p$ and $N^-(0) = 1-p$, where $2p-1$ is the initial value of $\varrho$. The solution to Eq.~\ref{rateeq} yields for the temporal dependence of the polarization: $\varrho (t) = (2p-1) \exp(-2t/t_{\text{iv}})$. Note that $t_{\text{iv}}$ is twice the polarization
decay time. The experimental data is fitted with this formula plus a constant background evaluated as a mean value of $\varrho$ for decays between $\sim$ 75 and 85 ps, accounting for the background polarization. This procedure allows to evaluate $p$ and $t_{\text{iv}}$. We find that for the exciton $p = 0.72 \pm 0.07$, approximately constant over the whole range of excitation fluences. For the trion, $p$ decreases monotonically from 0.75 to 0.65 as the fluence is increased. The fluence dependence of $t_{\text{iv}}$ plotted in Fig.~\ref{polar}(g) reveals that with increasing
fluence, intervalley scattering time decreases from $\simeq50$ ps to $\simeq20$~ps for both the exciton and trion.

A natural question arises whether the acceleration of the intervalley scattering can be attributed to the heating of the lattice by the laser pulse. Increased scattering via an increased acoustic phonon population is ruled out since it would require heating of the lattice to temperatures higher than 100 K.\cite{Zeng12} At the temperature of the experiments scattering on ionized impurities is expected to be the dominant momentum relaxation mechanism. An increased lattice temperature would also lead to a reduction of the band gap and, consequently, a larger $E_k$ since we keep the excitation energy constant. This would lead to a decrease of the initial polarization via an increase of both $\mathbf{\Omega}$ and $\tau_p$. However, as seen from the temporal traces of $\varrho$ presented in Figs.~\ref{polar}(a)-(f), polarization values at short time decays are approximately independent of the excitation fluence (a fact corroborated by a very weak dependence of the fitted $p$ as a function of fluence). Moreover, the dependence of the exciton transition energy on fluence (see Fig. \ref{trpl}(c)) demonstrates the absence of any lattice heating at least up to about 60 $\mu$J/cm$^2$.

Another mechanism leading to the acceleration of intervalley scattering could be related to a screening of the Coulomb potential of the ionized impurities. As shown for intrinsic GaAs quantum wells, this mechanism was found to be responsible for the decrease of spin relaxation time with increasing excitation density as a result of an increase in $\tau_p$.\cite{Munoz95} However, this mechanism would also affect the initial polarization and therefore can be ruled out in our case.

Our experimental results demonstrate that the increased intervalley scattering rate concerns only {\em cold} excitons, \emph{i. e.}, with a small center-of-mass momentum. We propose that the effect leading to the increased valley relaxation rate can be related to the heating of the exciton system. Indeed, larger average kinetic energy results in a larger $\mathbf{\Omega}$. The heating could occur as a result of an increased exciton-exciton scattering. Elevated quasi-particle temperature explains also the small decrease of the initial trion polarization; increased trion temperature leads to a small occupation of the excited trion configurations\cite{Jones13,Yu14d} thus opening additional intervalley scattering channels. As shown by the temperature dependent time-resolved Kerr rotation studies, enhanced intervalley relaxation requires exciton temperatures higher than 30~K.\cite{Zhu14} We remark that at
such high temperatures we expect a significant increase of the exciton radiative lifetime, in agreement with a recent theoretical work.\cite{Palummo15} However, the measured $t_1$ does not exhibit any dependence on fluence suggesting that $t_1$ contains a significant contribution from non-radiative processes.

Another possible origin of the increased intervalley scattering rate can be attributed to the presence of a dense exciton gas. We calculate the fluence dependent average exciton-exciton distance by assuming that about 1\% of the photons are absorbed by the WSe$_{2}$ monolayer\cite{Li14} and that each absorbed photon is converted into an exciton. This crude assumption gives a lower (upper) bound for the average exciton-exciton distance (exciton density). The calculated values are given in the upper axis of Fig. \ref{polar}(g). It shows that at the highest fluence the average
exciton-exciton distance becomes comparable to the exciton Bohr radius of 1.3 nm.\cite{Srivastava14}. Under such conditions, screening and/or many body effects are expected to influence the exciton valley dynamics.

In conclusion, we have studied the exciton and trion PL in monolayer WSe$_{2}$. The intensity of the exciton PL shows no signs of saturation up to the highest excitation densities where heating effects are already present. This demonstrates that non-linear processes such as exciton-exciton annihilation or biexciton formation are inefficient at low temperatures and for quasi-resonant excitation. In addition, we have shown that dark non-radiative trap states play an important role in determining the PL decay dynamics. However, at high excitation densities these states become saturated. Finally, the measured intervalley scattering rates of both excitons and trions are significantly increased with increasing excitation density which may have negative effects upon the operation of \emph{valleytronic} devices. Importantly, the increased intervalley scattering rate occurs only for cold excitons and does not influence hot excitons -- created with a high kinetic energy. We attribute this effect to either heating of the exciton system by increased exciton-exciton scattering or to the presence of a dense exciton gas leading to a modification of the exciton valley properties.

This work was partially supported by  Programme Investissements d'Avenir under the program ANR-11-IDEX-0002-02 - reference ANR-10-LABX-0037-NEXT, ANR JCJC project milliPICS, the Region Midi-Pyr\'en\'ees under contract MESR 13053031 and STCU project 5809. ŁK acknowledges the support of Institut National des Sciences Appliqu\'{e}es Toulouse.


%

\end{document}